\documentclass[runningheads]{llncs}

\usepackage{graphicx}

\usepackage{tikz}
\usepackage{comment}
\usepackage{amsmath,amssymb} 
\usepackage{color}
\usepackage{mathtools}

\begin{document}

\pagestyle{headings}
\mainmatter
\def\ECCVSubNumber{100}  

\title{Block-matching in FPGA} 

\titlerunning{Block-matching in FPGA}
%
\author{Rafael Ulises Luzius Pizarro Solar \and
Michal Pleskowicz}
\authorrunning{R. Pizarro and M. Pleskowicz}
\institute{EPFL, Lausanne, Switzerland}

\maketitle

\begin{abstract}
Block-matching and 3D filtering (BM3D) is an image denoising algorithm that works in two similar steps. Both of these steps need to perform grouping by block-matching. We implement the block-matching in an FPGA, leveraging its ability to perform parallel computations. Our goal is to enable other researchers to use our solution in the future for real-time video denoising in video cameras that use FPGAs (such as the AXIOM Beta).
\end{abstract}
\vfill

\hfill{\scriptsize This project was supervised by Majed El Helou in CS413 at EPFL}
\newpage
\section{Intro}
Ever since the conception of photography there was noise present in the pictures - first in the form of film grain, that is a random optical texture due to the presence of metallic silver particles, and later on, when digital photography arrived, the noise became a mix of sensor's shot, thermal and read noise, among others. In general one image's pixel values can be represented as $$y = x + \alpha $$ where x is the natural noise-free image, y is the given resulting image and $\alpha$ is usually approximately assumed to represent White Additive Gaussian Noise; in color images that equation is applied across all three channels.
\par The goal of image denoising is to remove as much noise from the given image as possible, without losing the original image features (such as edges and textures).
\par Digital photography introduced the ability to perform image transformations more complex than the ones used in film labs back in the day. Various image denoising techniques have been proposed since then: from simple spatial filtering to recent developments in Convolutional Neural Networks (CNN). We'll briefly summarize two algorithms of the well-performing image denoising techniques with their respective limitations for an FPGA implementation, and then the one we settled on: BM3D.
\bigskip

\section{Related literature}
\subsection{Bilateral filtering}
Bilateral filtering is a non-linear extension of the simple Gaussian smoothing. The Gaussian smoothing applies the same kernel across all the pixels of an image, resulting in an output with a lower noise. To use it, we perform a convolution of the noisy input image with a Gaussian 2D function, which is defined as: 

$$G_\sigma(x)=\frac {1}{2\pi \sigma ^{2}}e^{-{\frac {x^2}{2\sigma ^2}}}$$
\par Since the Fourier transform of a Gaussian function is another Gaussian function, the convolution with a Gaussian kernel is essentially a low-pass filter, meaning a filter that removes high frequencies in an image. In addition to removed noise, images produced by this operation tend to have blurred edges all across the image.
\par Bilateral filtering mitigates this problem by introducing non-linearity. In it's essence, bilateral filtering also constitutes of a weighted average of the pixel values in the input pixel's neighbourhood. However, in comparison to the Gaussian smoothing, it takes into consideration the difference in value between the input pixel and pixels in its neighbourhood when assigning weights for the subsequent averaging. The filter is defined as:
$$BF[I]_p={\frac {1}{W_{p}}} \sum_{q \in S} G_{\sigma_s} (||p-q||) G_{\sigma_r} (I_p - I_q) I_q,$$
where the normalization factor $W_p$ is:
$$W_p=\sum_{q \in S} G_{\sigma_s} (||p-q||) G_{\sigma_r} (I_p - I_q)$$
It is made of two 2D Gaussian functions. The $G_{\sigma_s}$ is the spatial Gaussian, that decreases the weight of the pixels further away spatially from $I_p$, and the $G_{\sigma_r}$ is the range Gaussian which decreases weights of pixels that have a higher difference in intensity when compared with the input pixel $I_p$.
\par A limitation of the Bilateral filtering is that it often smoothens the textures as well as produces a 'staircase' effect in larger spots with color gradients.

\subsection{Fast and Flexible Discriminative CNN Denoiser (FFDNet)}
FFDNet \cite{ffdnet} was conceived in order to tackle the main issues with discriminative learning methods used for image denoising, like DnCNN \cite{dncnn}. The neural network is formulated as 
$$ x = \mathcal{F}(y,M;\Theta) $$
where:
\begin{itemize}
    \item $y$ is the input image
    \item $M$ is a tunable noise level map
    \item $\Theta$ denotes the model's parameters
\end{itemize}

\begin{figure}[h]
    \centering
    \includegraphics[width=0.7\linewidth]{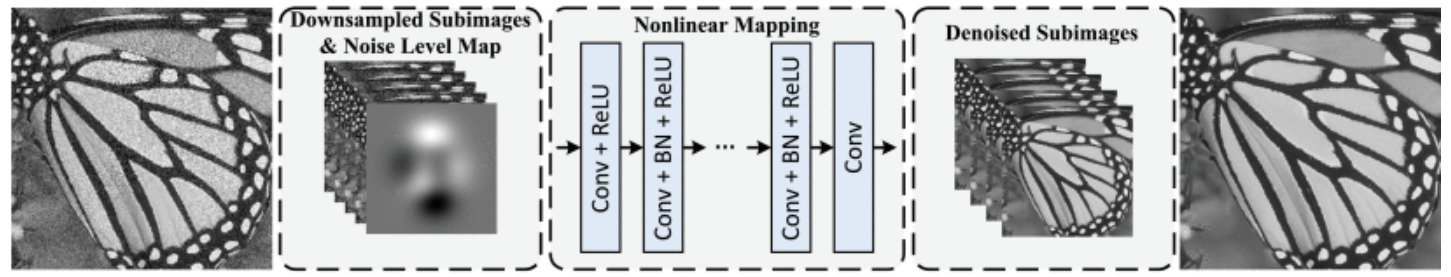}
    \caption{FFD Net structure | source:\cite{ffdnet}}
    \label{ffd}
\end{figure}

The network's structure can be seen in the Figure \ref{ffd} \newline
The first layer is a reversible downsampling operator, that reshapes the input $x$ into four downsampled sub-images.
\par The following CNN consists of a series of 3x3 convolution layers. Zero-padding is used in all of them in order to keep a constant size of the feature maps.
\par Following the last convolution layer, the reverse of the downsampling operator used in the first layer is applied. The number of convolutional layers was set to 15 for greyscale images and 12 for color images.
\newline FFDNet promises to:
\begin{itemize}
    \item handle a wide range of noise levels ($\sigma \in [0,75]$) with a single network
    \item remove spatially invariant noise by applying the non-uniform noise map $M$
    \item achieve faster speed than benchmark BM3D without sacrificing performance
\end{itemize}
\par Recent improvements to this approach exist, most notably BUIFD \cite{ffd_improved} that predicts the noise level map within the network. 
\par In general, the neural net model is too big to be fitted in an FPGA of our target size (Apertus Project \cite{Apertus}).

\section{Block-matching and 3D filtering (BM3D)}
\begin{figure}[h]
    \centering
    \includegraphics[width=0.9\linewidth]{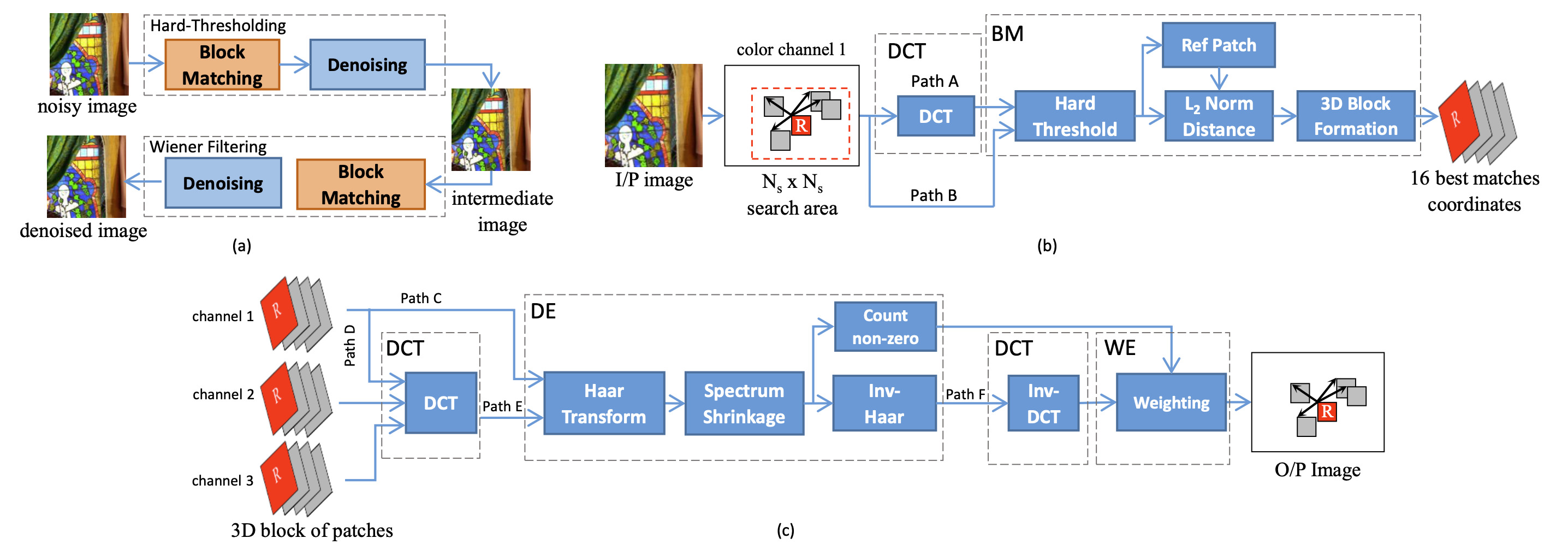}
    \caption{BM3D algorithm pipeline. On the top left we can see the two major steps: Hard-thresholding, that outputs the intermediate image and Wiener filtering, that gives the final output. On the top right is the block-matching (BM) that extracts similar blocks' coordinates from a single channel. On the bottom there is the reduction from similar patches list to the denoised image.   |  source: IDEAL \cite{ideal}}
    \label{bm3dstructure}
\end{figure}

First proposed by Dabov et al. in 2006 \cite{bm3d}, BM3D denoises the image by finding similar patches inside it and leveraging their mutual correlations. The algorithm works in two steps: Hard-thresholding and Wiener filtering, which can be seen in Figure \ref{bm3dstructure}.
Both of them are constructed from these substeps: \newline
$$ \text{Block-matching} \to \text{Collaborative Filtering }\to \text{Aggregation} $$
\newline We picked Marc Lebrun's \cite{lebrun} as the reference open-source implementation of a working full BM3D algorithm. All the denoising is performed on the luminance channel of the input transformed to YCbCr color space, as the luminance channel contains enough information of block similarity.

\subsection{Hard-thresholding}
\par Since the parameter names repeat in the two steps we add a $hard$ subscript to them in this section 
\par We denote by $P$ the reference patch of size $k^{hard} \times k^{hard}$. Define a searching window of size $n^{hard} \times n^{hard}$, which is the neighbourhood in which the algorithm looks for patches similar to the reference one (the searching window being centered on P). For every possible patch in the input we look for possible candidates for the set of similar patches, which is defined by

$$\mathcal{P}(P) = \{Q : d(P,Q)\leq \tau^{hard} \} $$

where: 
\begin{itemize}
    \item $\tau^{hard}$ is the distance threshold for d, under which two patches are considered to be similar
    \item $d(P,Q) = \frac{||{\gamma ' (P) - \gamma ' (Q)|| ^2 _2}}{(k^hard)^2}$ is the normalized quadratic distance between the patches
    \item $\gamma '$ is a hard-thresholding operator with threshold $\lambda _{2D} ^{hard} \sigma$
    \item $\sigma ^2$ is the assumed variance of the white additive Gaussian noise
\end{itemize}

\par All the possible patches in the searching window are compared with the reference patch, and the $N^{hard}$ closest results are kept as the stacked 3D group denoted by $\mathbb{P}(P)$. Only the positions of the patches in the image are stored in the stack, not the pixels values that they contain.

\par The next substep is the collaborative filtering. The algorithm applies a 3D isometric linear transform to the whole group, then shrinks the transform spectrum. After that, the inverse linear transform is applied. For each patch we obtain
$$ \mathbb{P}(P)^{hard} = \tau _{2D} ^{hard^{-1}} (\lambda(\tau _{2D} ^{hard}(\mathbb{P}(P)))) $$

The last substep is aggregation, which provides estimates for each used patch. These give us a variable number of estimates for every pixel
$$ \forall Q \in \mathcal{P}(P), \forall x \in Q,
    \begin{cases}
      \nu (x) = \nu (x) + w ^{hard} _{P} u^{hard} _{Q,P}(x)\\
      \delta (x) = \delta (x) + w ^{hard} _{P}
    \end{cases} $$
where:
\begin{itemize}
    \item $v$ and $\delta$ denote the numerator and the denominator parts of the final estimate
    \item $u^{hard} _{Q,P}(x)$ denotes the estimate of the pixel $x$ in patch \( \mathcal{Q} \)obtained during collaborative filtering of patch  \( \mathcal{P} \)
    \item  $w ^{hard} _{P} = \begin{cases}
      (N ^{hard} _P)^{-1} & \text{if $N ^{hard} _P \geq 1$}\\
      1 & \text{otherwise}
    \end{cases}$
    \item $N ^{hard} _P$ denoting the number of valid coefficients in the 3D block after hard-thresholding
\end{itemize}
This weighting scheme prioritizes homogeneous patches, which lets it avoid artifacts around the edges of the input image. A Kaiser window of size $k^{hard} \times k^{hard}$ (patch size) is used as a part of the weights. We simply multiply it with the result of the inverse linear transform. 
The basic estimate for this step is given by 
$$ u^{basic} (x) = \frac
{\sum_{P} w ^{hard} _P \sum_{Q \in \mathcal{P}(P)} X_{Q}(x)u^{hard} _{Q,P}(x)}
{\sum_{P} w ^{hard} _P \sum_{Q \in \mathcal{P}(P)} X_{Q}(x)} $$
where:
\begin{itemize}
    \item $X_{Q}(x) = 1 \iff x \in Q$, 0 otherwise 
\end{itemize}
It is the division of the nominator and the denominator mention before, per element.

\subsection{Wiener filtering}
The second step makes use of the basic estimate $u^{basic} (x)$ calculated above. It filters the input image $u$ with the Wiener filter, but uses the $u^{basic} (x)$ as a reference point. This second step has been shown in experiments to restore more detail in the image and to improve the denoising performance.

Once again we look for the set of similar patches, this time denoted by:
$$\mathcal{P}^{basic}(P) = \{Q : d(P,Q)\leq \tau^{wien} \} $$
After that set has been obtained, we form two 3D groups:
\begin{itemize}
    \item $\mathbb{P}^{basic}(P)$ by stacking up patches from the basic estimation $u^{basic} (x)$
    \item $\mathbb{P}(P)$ by stacking up patches in the same order from the original noisy image $u$
\end{itemize}
We only keep $N^{wien}$ best patches in each of them. 
\par Collaborative filtering is next applied on these two 3D stacks. We define empierical Wiener coefficients as follows:
$$ \omega _{P} (\xi) = \frac{|\tau _{3D} ^{wien} (\mathbb{P}^{basic}(P)) (\xi)|^2} {|\tau _{3D} ^{wien} (\mathbb{P}^{basic}(P)) (\xi)|^2 + \sigma ^{2} } $$
The filtered 3D stacks are then obtained by 
$$ \mathbb{P}^{wien}(P) = \tau _{3D} ^{wien^{-1}} (\omega _P \cdot \tau _{3D} ^{wien} (\mathbb{P}(P))$$
Then the foloowing buffers are created, just like in the same equation in Hard-thresholding
$$ \forall Q \in \mathcal{P}(P), \forall x \in Q,
    \begin{cases}
      \nu (x) = \nu (x) + w ^{wien} _{P} u^{wien} _{Q,P}(x)\\
      \delta (x) = \delta (x) + w ^{wien} _{P}
    \end{cases} $$
where $w ^{wien} _{P} = ||\omega _{P}|| _2 ^{-2}$
\par Similarily, the Kaiser window of size $k^{wien} \times k^{wien}$ is applied to reduce border effects. The final estimate $ u^{final} (x)$ is given by
$$ u^{final} (x) = \frac
{\sum_{P} w ^{wien} _P \sum_{Q \in \mathcal{P}(P)} X_{Q}(x)u^{wien} _{Q,P}(x)}
{\sum_{P} w ^{wien} _P \sum_{Q \in \mathcal{P}(P)} X_{Q}(x)} $$
where:
\begin{itemize}
    \item $X_{Q}(x) = 1 \iff x \in Q$, 0 otherwise 
\end{itemize}

\section{Implementation}

While GPUs are able to expose high amounts of parallelism for vector based algorithms, they are power hungry (hundred of watts), and once the algorithm gets further away from wide vector operations the overall GPU performance drops significantly. As an example, a GPU implementations of BM3D can spend up to 87\% of time in the block-matching step \cite{ideal} showing that BM3D mismatches with the GPU architecture.

The reconfigurability of FPGAs makes them a perfect candidate for real-time image processing because all of the parallelism available can be exposed to hardware given enough resources. FPGAs are also interesting from the camera building perspective as features can be added over time, or a specific algorithm can reprogrammed and executed depending on the context. The Apertus project \cite{Apertus} is an example of an open camera with an FPGA as the base compute unit, so any potential image processing FPGA architecture could fit integrated in their pipeline.

ASIC would be a step further, but ASIC affordability comes with volume, custom architectures are usually part of niche market and can be ill-fitting for ASIC design. In addition to that, ASIC design is more time-consuming and expensive than working with FPGAs.

In this project, we implement an architecture targeting Xilinx FPGAs, although the principles exposed could be integrated for any kind of custom accelerator.

\subsection{Architectures}

We can approach the block extraction step from two architectural perspectives, per Block or per Stream.

\subsubsection{Block}
The usual way of approaching the BM3D algorithm is from a block perspective. Our goal is to search for similar blocks within a defined window size, so the natural intuition would be to create a unit that stores the block (bS), store the window size (wS), and then iterate over every pair within the window size. We can see the execution flow in Fig. 3: we compute pair by pair the sum of difference between blue square and the reference black square, where  the green square represents a block later in time.

\begin{figure}[h]
    \centering
    \includegraphics[width=0.8\linewidth]{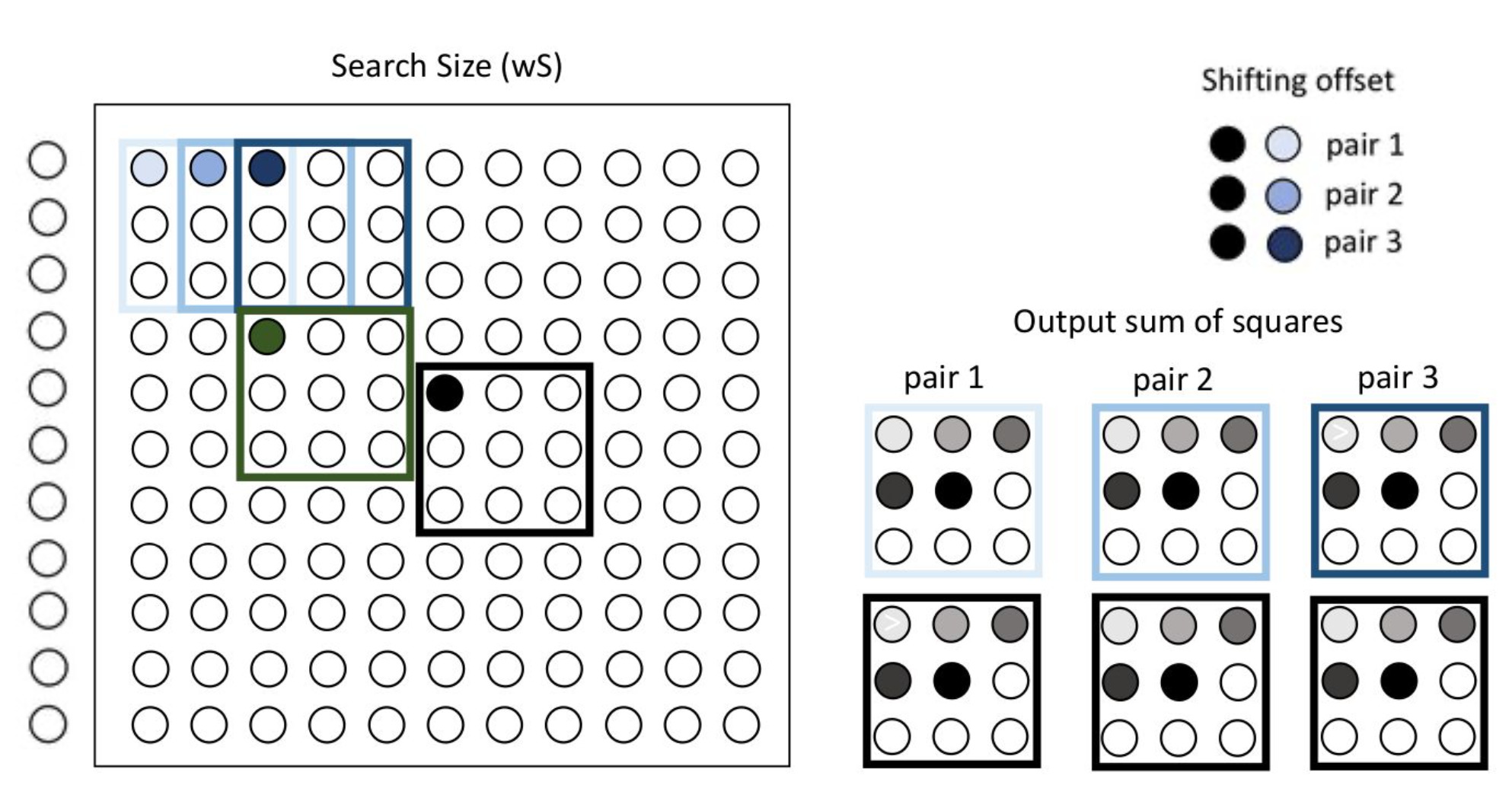}
    \caption{Shows execution flow of a block worker within a search window frame. The reference block is in black. We calculate the distance between two blocks by comparing pixel by pixel the reference block to every possible block within the Search Window. The blue colors represent the shifting offset selection. For each offset we perform a full differentiation of the block size (here bS*bS = 3x3), so it takes 9 cycles to calculate distance between two blocks.}
    \label{blockwise}
\end{figure}

In terms of resource requirements, each worker should have : 
\begin{itemize}
    \item buffer space to store the full search window (wS*wS)
    \item buffer space to store the reference block    (bS*bS)
    \item a memory controller to read pixel per pixel
    \item buffer space to store best N results
\end{itemize}

As an example, a search window size of 32 would fit in a single BRAM resource, and a reference block of size 8 and smaller should be mapped to LUT registers.
Note that memory access patterns are 2d blocks, and memory performs best when reading row per row, so filling the search window can yield sub-optimal DRAM throughput \cite{memory}.

\subsubsection{Stream}
\begin{figure}[h]
    \centering
    \includegraphics[width=0.8\linewidth]{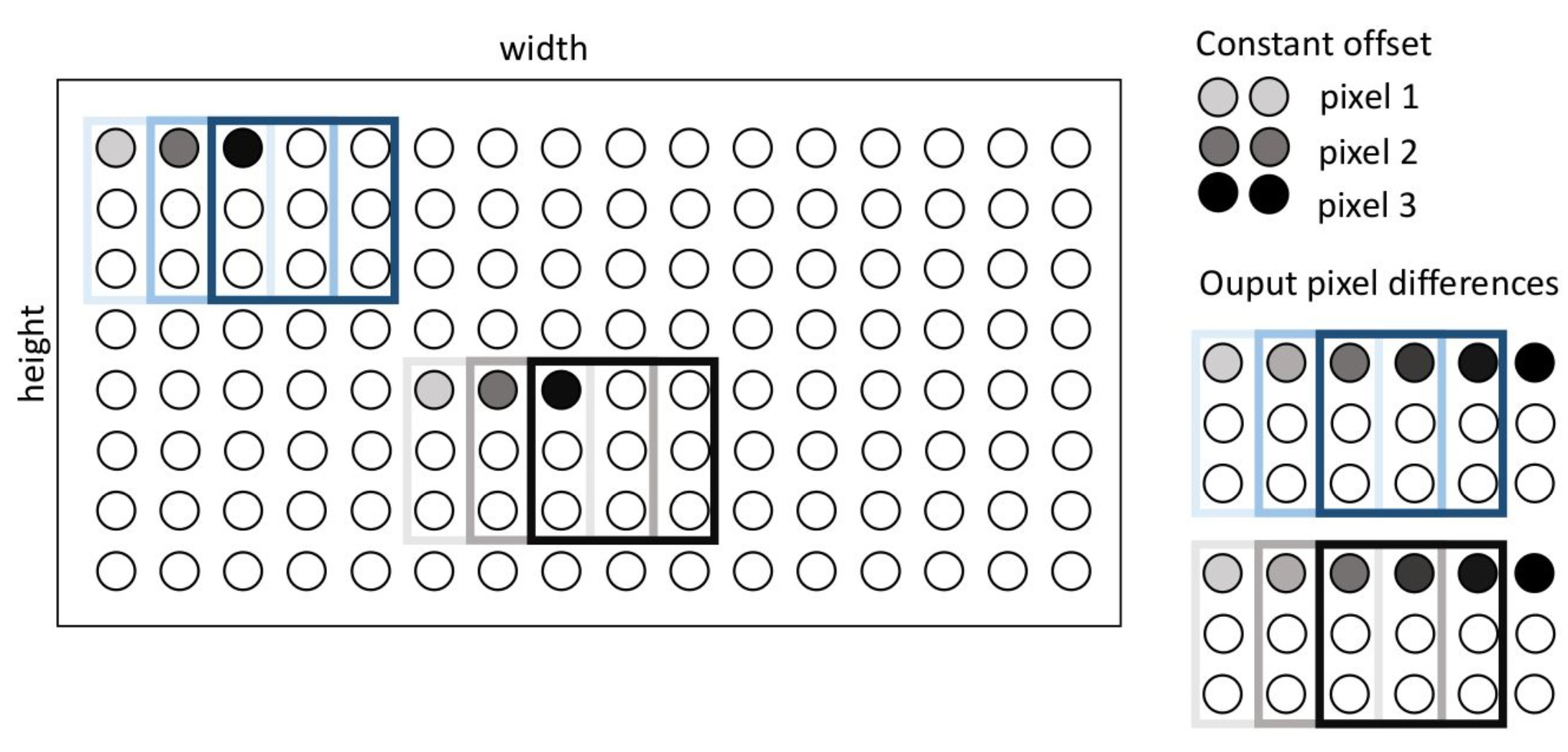}
    \caption{Shows the execution flow generating difference with given offset from input image. In black we can see the sliding reference block, and in blue the sliding base block. Instead of calculating every pair for a given reference block like in Fig. \ref{blockwise}, we calculate a single pair for a reference block given that offset. For that same offset, every other reference block in the frame also calculates the distance with its pair block.}
    \label{streamwise}
\end{figure}

A more novel approach is to see that if we continue sliding, we can calculate the difference of the next reference block, instead of stopping the flow and jumping to next row of the reference block, Fig. \ref{streamwise}. shows us the intuition behind this. So we can shift our perspective from a single block and it's window size, to see the problem as generating the differences of a given offset between the whole image, then aggregating them for every offset possible. In this approach, the full image becomes the reference block and base block, a single differential image corresponds to a specific offset for the reference block in the search window, and given a single differential image, we can only calculate one pair distance of a specific block. We can see the execution flow in Fig. \ref{streamwise}, note that the offset between pairs is static, and that it'll take 3 full rows in this example before having all differences necessary for a full sum of the initial shown blocks.

In terms of resource requirements : 
\begin{itemize}
    \item initially buffer enough rows to have both the offset pixel and input pixel available in the FPGA Memory $$O(wS*width)$$
    \item buffer rows for deeper pipeline steps depending on $$O(bS*width)$$
\end{itemize}

\subsection{Implementation}

We choose the Stream-based approach, as we believe that full row-based memory access patterns can yield maximum DRAM throughput and the Block-based approach doesn't make good use of neighbour locality, and finally we want to explore this new design space, as the Block-based approach has been done previously in FPGA \cite{cardoso} and in ASIC \cite{ideal}.

We assume that the image arrives as a continuous stream, although our implementation supports granularity per continuous row. We also advise users to perform a transpose on the input, as this will significantly reduce the resource utilization for wider pictures.
The pipeline is decomposed in the following steps:
\begin{itemize}
    \item From Image to Square Difference
    \item From Square Difference to Sum
    \item Take Best N Blocks for Reference Block
\end{itemize}

\subsubsection{From Image to Square Difference}

\begin{figure}[h]
    \centering
    \includegraphics[width=0.7\linewidth]{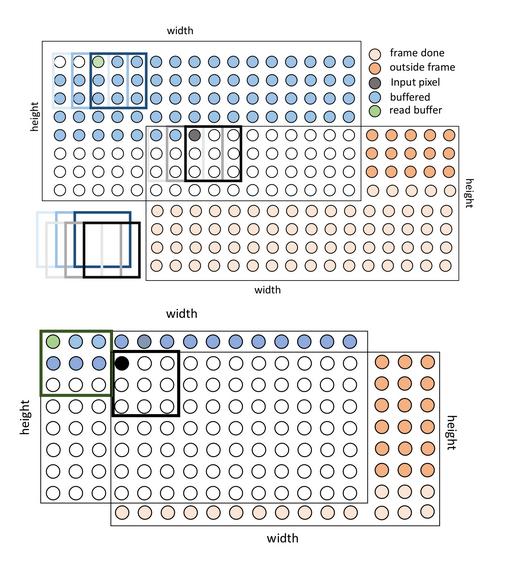}
    \caption{Shows the execution flow of generating the differential image with two different offsets. The top offset is 4 rows and 5 columns, bottom one is 1 row and 3 columns. We can see that the part of the image in red is where the differentiating pixels are outside the frame. In blue is the buffer space. The Black Boxes are the reference blocks, their base coordinates are from the top left frame, the bottom right frame represents an imaginary sliding window of the black box. }
    \label{streamchallenge}
\end{figure}

The first step of our design is to implement the Square Difference from an image. This step consists of calculating the difference between every pair of two pixels with a given offset from a continuously streaming input image stream. On the top part of image of Fig \ref{streamchallenge} we can see the sliding blocks. During each cycle we calculate the squared difference of the green buffered pixel and black input image pixel. We need to buffer (wS*width/2) elements to point to the green pixel. Every iteration lowers the amount of buffered pixels in order to shift its offset through the search window.

Here we note that the differential image varies in size depending on the offset. Let's imagine (1) an offset right next to the reference and another one right at the (2) edge of the search window. (1) Only the pixels on the right most edge won't have a pixel at that offset, making the image size (width-1)*height. (2) Pixels within a half window size (wS) won't have pixels at that offset, making the image size width-wS. This is illustrated by the difference in outside frame pixels between the two images in Fig \ref{streamchallenge}. To overcome this, we choose to only stream the set of pixels where all the differential images overlap.

\subsubsection{From Square Difference to Sums}

\begin{figure}[h]
    \centering
    \includegraphics[width=0.7\linewidth]{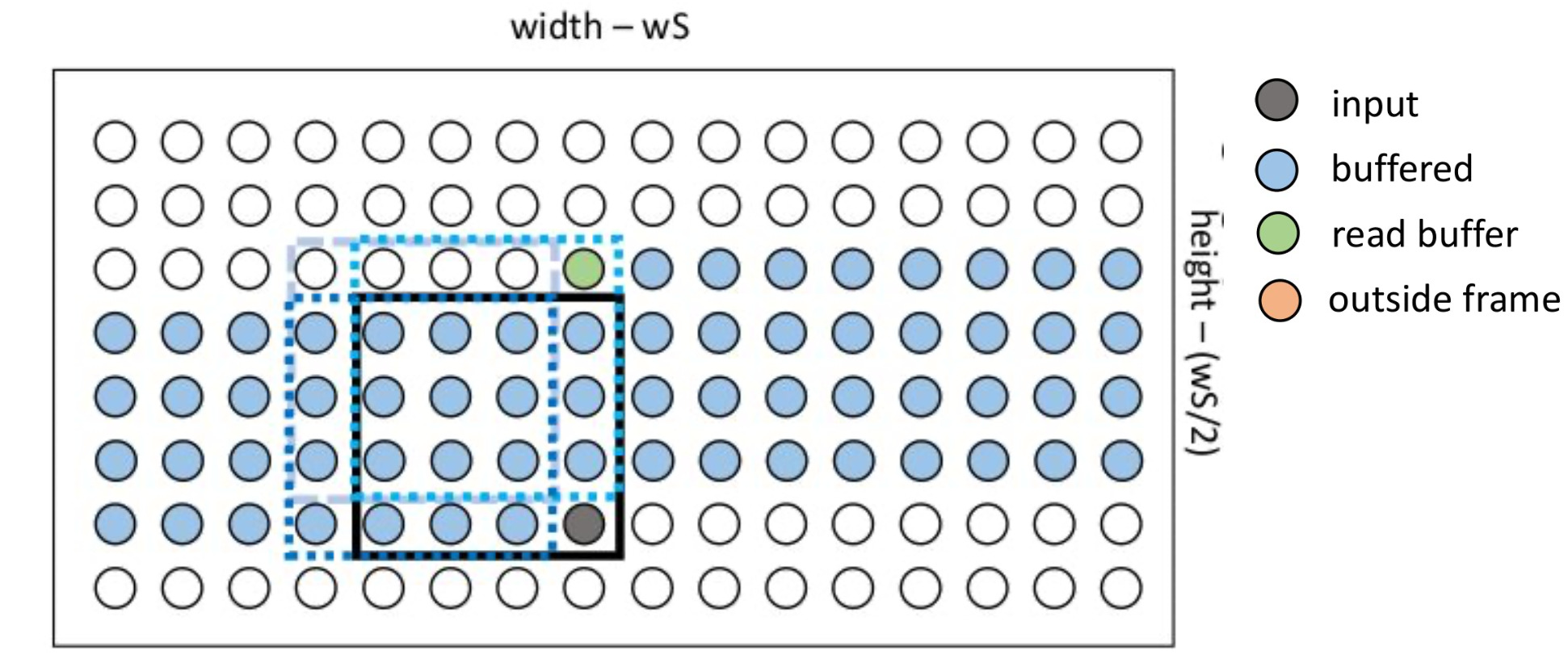}
    \caption{The sliding compute block receiving differential pixel as input stream and output streaming total sum from previously buffered elements. A more detailed figure of the compute block is present in Fig.\ref{computeblockdetail}. The top right pixel is consumed from the pixel buffer, the top blue square is also consumed from the sum buffer, other elements are available by shift registers. }
    \label{blockslide}
\end{figure}

Now that we have an input stream of differential pixels, we can calculate the total sum by using a sliding computing block. In Fig \ref{blockslide} we can see the sliding block computing the total black block sum from the previous computed values. Note that the amount of buffering necessary to compensate for top right pixel is width*bS, and the amount of buffering necessary for the past row sums is a single row.

\begin{figure}[h]
    \centering
    \includegraphics[width=0.5\linewidth]{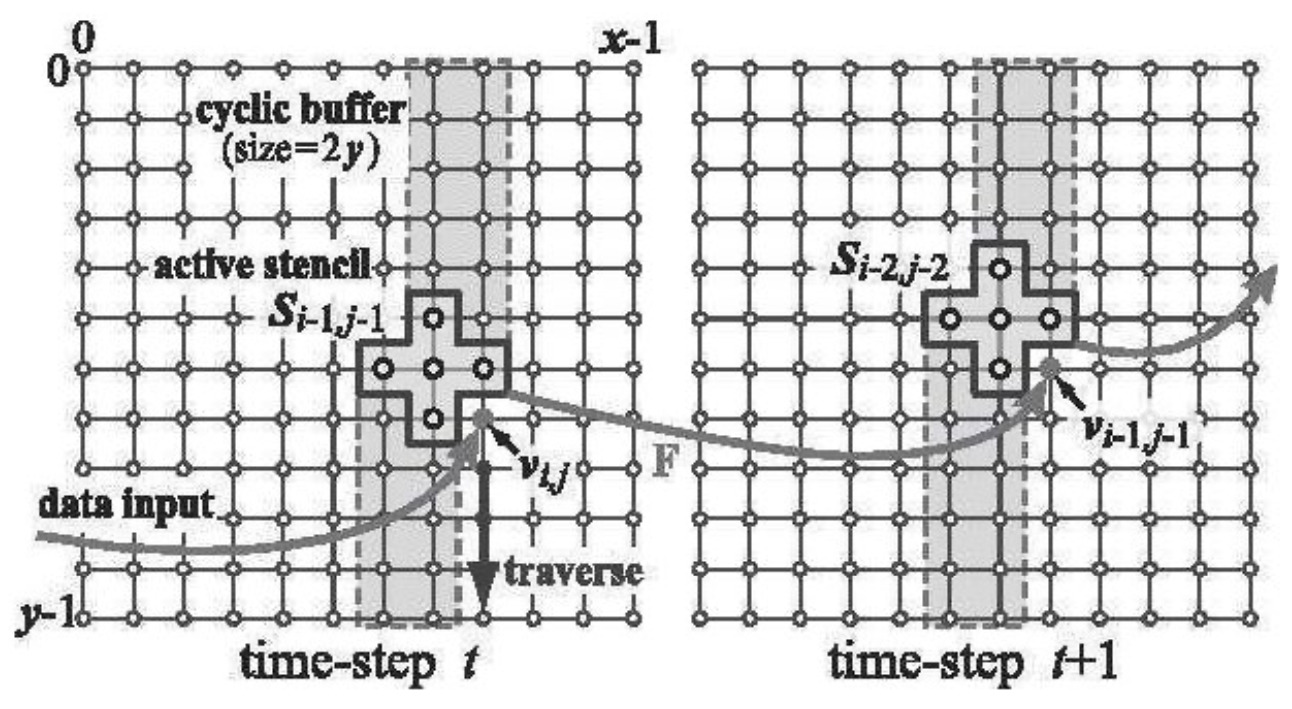}
    \caption{Classical stencil block sliding. In grey we can see the buffered elements. The next element is calculated from current block, consuming buffered element and current input.}
    \label{stencil}
\end{figure}

Compute blocks sliding while buffering intermediate values is known as the stencil based algorithms \cite{stencil}. Algorithms with this form are the perfect examples that can be implemented in FPGAs (Fig \ref{stencil}).

\subsubsection{Compute Block}

\begin{figure}[h]
    \centering
    \includegraphics[width=0.6\linewidth]{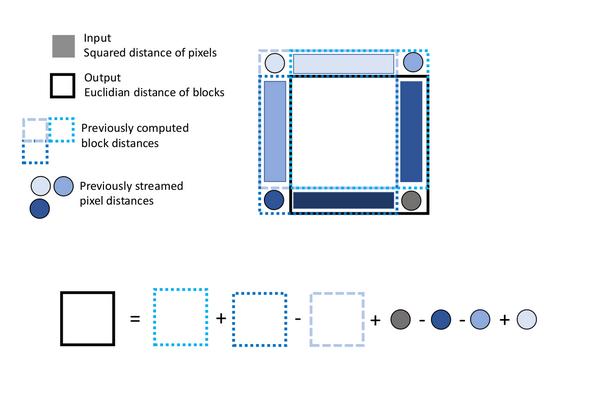}
    \caption{Details of the sliding compute block. Squares are total sums, circles are pixels. In blue are previously computed elements, black is the current reference patch. The black square can be seen as the addition of previously computed elements.}
    \label{computeblockdetail}
\end{figure}

The compute block in Fig \ref{computeblockdetail} shows us the data dependencies between cycles. Here we can see the dependencies between elements, and how the current sum (black) is generated for previous values (blueish) and current input (black pixel). The block itself is based on a well known trick in image processing: the Summed area table \cite{summedarea}.

\begin{figure}[h]
    \centering
    \includegraphics[width=0.3\linewidth]{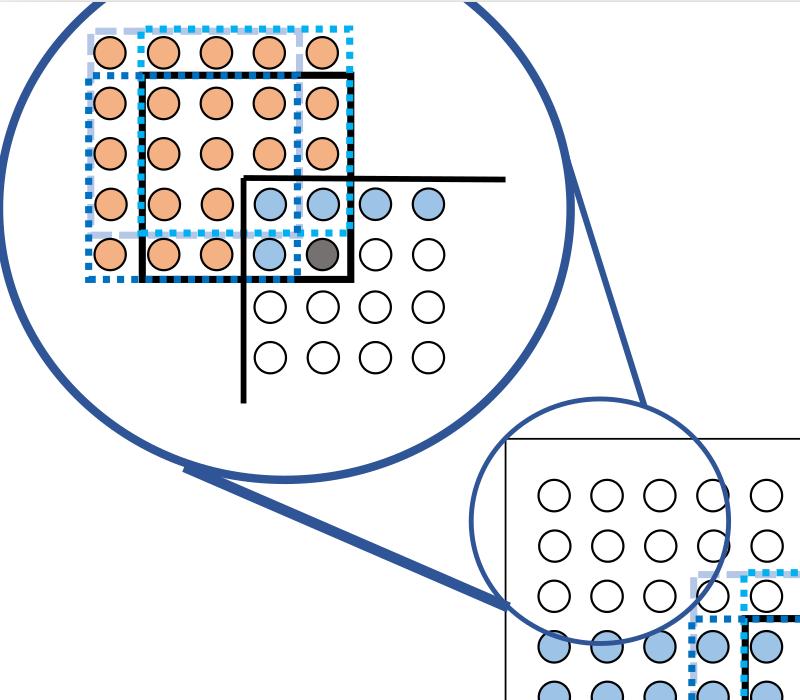}
    \caption{Details of sum when not completely in frame. Red pixels count as zero. Here the left sum is the sum of two pixels to the left, where both top and bottom left pixel corners are zero.}
    \label{blockchallenge}
\end{figure}

One challenge to consider is when the block is on the edges of the image as shown in Fig \ref{blockchallenge}. To compensate for this, elements which are not yet available are valued at zero, e.g. the top sum table before the completion of the first row, and the left pixels before column offset of patch size.

\subsubsection{Optimizing}

\begin{figure}[h]
    \centering
    \includegraphics[width=0.8\linewidth]{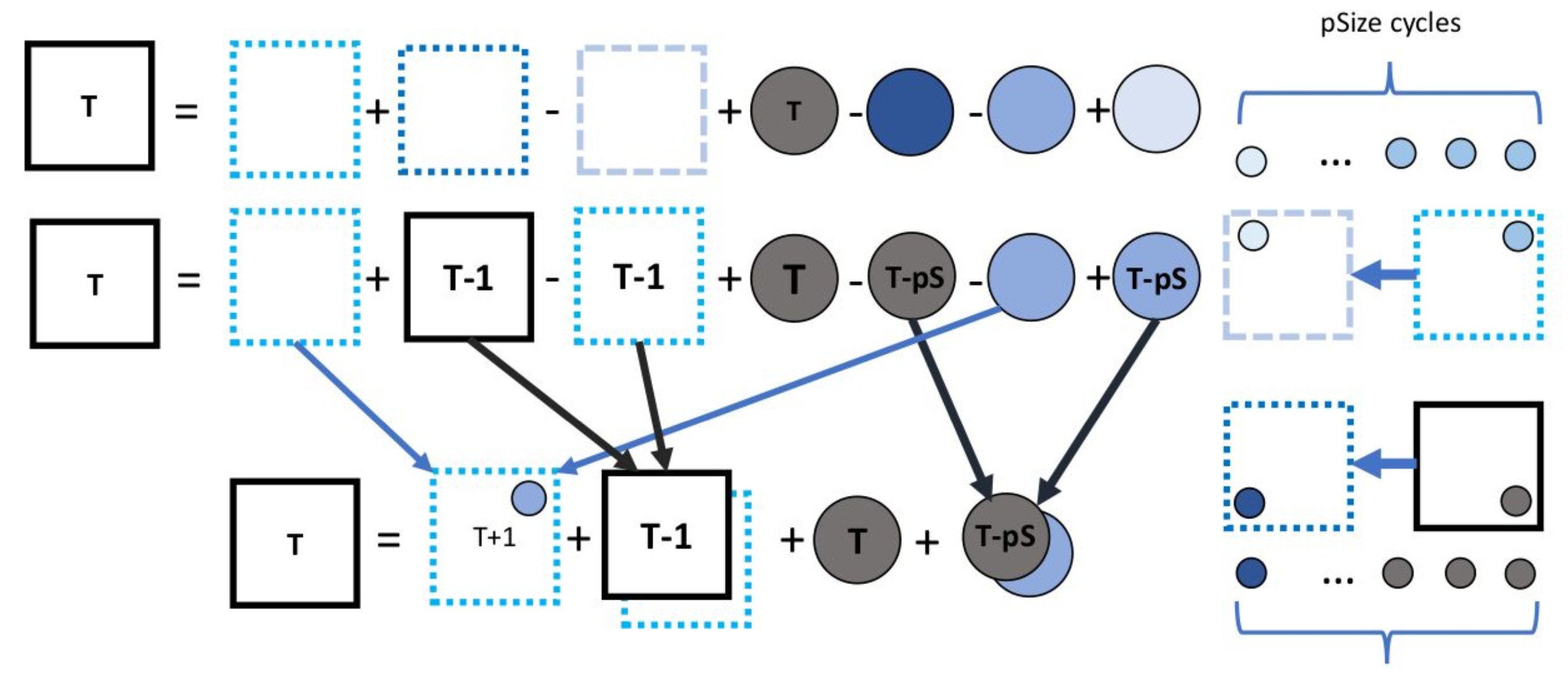}
    \caption{Optimizing by showing data dependencies. First row represents the non-optimized data dependencies. Second row represents exposing temporal dependencies, where one element at a given cycle becomes another element later in time. Third row arrows represent computations done in cycles before actually calculating the total sum at time T. Right figure shows data dependencies as one slide from one sum to the next.}
    \label{pipelining}
\end{figure}

As we are now able to generate the total sum from an input image, we should analyse where we stand in terms of performance and resource utilization. After synthesis using Vivado tools we achieve a little over 100MHz without optimization, but fail to meet timing on 250MHz. Analysing the timing report we detect that we have around 16\% of endpoints failing with a Worst Negative Slack (WNS) of -2ns. Selecting for the paths that contain the worst slack we can see that the sum of our compute block is the longest path. This is to be expected, given that the current amount of additions of 32 and 18 bits to do in a single cycle is 6. While pipelining is the usual trick implemented in case of increasing an operation performance, it cannot apply when T depends on T-1 unless you divide your throughput by the number of pipeline stages. To overcome this issue we expose other kinds of dependencies seen in Fig \ref{pipelining} such as using temporal delays both by prereading (T+1) and precomputing (T+1->T, T->T-1, T->T-pS) values. We could try to push this concept further and get rid of 1 extra addition by prereading even deeper, but this increases the cost in resources and in synchronization.

\subsubsection{Parallelization}

\begin{figure}[h]
    \centering
    \includegraphics[width=0.7\linewidth]{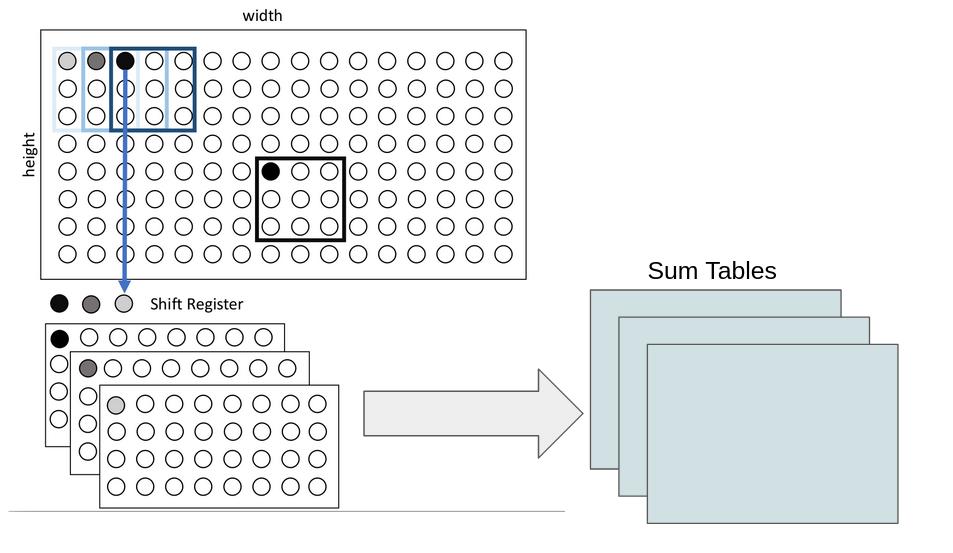}
    \caption{Parallelization by shifting the input pixel. The initial pixel of each worker shows that it's calculating different offsets. In blue we can see the initial difference square to calculate for 3 different workers. This would yield 3 sum tables, each one representing an unique offset, and each first element of a sum table would correspond to the same reference block.}
    \label{parallelize}
\end{figure}

Now that we can compute, from a single image stream, the total sums between every pair of blocks of a given offset, we need to see how to squeeze more throughput by parallelizing. Each offset result is independent from each other, which means we could calculate various offsets in parallel. This is done by shifting the image input on the Image to Square phase by the number of workers as shown in Fig \ref{parallelize}. On each image input, we'll fully execute the matching number offset to workers, so for the next image iteration we increase the offset of the base pixel reference by the number of workers. At the last offset of a row in the search window, we jump to the initial offset of next row. This trick effectively multiplies the throughput by the number of workers.

To simplify the architecture, use full potential throughput, and use less resources we restraint the number of workers argument to be a multiple of the window width size :

$wSize \mod nbWorkers \equiv 0 $

This parallelization method can only be executed row-wise in the search window, as shifting the input would have no meaning beyond the search window size. If more than the search window row size of workers fit in the FPGA, we could modify in the future the design to parallelize row wise, simply by passing as an argument an initial row offset to a worker.
In any case, parametrizing for more than half the window size of workers would be very aggressive in terms of throughput and can potentially stall further down the pipeline.

\subsubsection{Pick N Best}

While our actual implementation stops at the previous step, here we give architectural recommendations to achieve the full high performing block-matching system.
Pick N best out of a stream is a well defined algorithm. The optimal solutions always takes log(N) elements to insert, making it an algorithm given M elements and N best of complexity O(M*log(N)).
This log(N) factor must be taken into account, as we'll need to reduce our throughput to match it. One way to do this without stalling the pipeline is by implementing a stride between blocks. Striding sums has shown to yield close to identical results in the reference BM3D implementation \cite{lebrun} and has been used in previous accelerators \cite{ideal}, \cite{cardoso}. In our case, striding is implemented fairly easy by only validating the sum output every stride clocks, thus decreasing throughput by a factor of stride and reducing DDR memory requirements by the same factor. Although this reduces the total of compared blocks by a factor of stride, it has been shown to not lower the final image quality \cite{lebrun}, so we can assume the throughput stays maximal.
A final factor to take into consideration is that the sum image's indices are not linear in memory, e.g. the search window of the first reference block is contained in the first entry of every sum table. This will require some smarter memory management such as a butterfly interconnect or similarly to a realigning pipeline to serialize data per reference block for the Pick N Best step, both which can run pipelined at high frequencies impacting the throughput very little.

\subsection{Final Note}
After further review, we could easily eliminate the Image to Square stage buffer. This can be done by reading both in parallel the image from the base value and from the current window offset. This would get rid of the blue pixel buffer and would stream both the green pixel and black pixel referenced in Fig \ref{streamwise}. It would have as a side effect more pressure on DDR, which shouldn't matter as row DDR access can throughput up to 533MHz (double our frequency). And it would complexify by a small margin the memory management, as right now there's none required.

\section{Results}

\subsection{BM3D}
\begin{figure}[!htb]
\minipage{0.32\textwidth}
  \includegraphics[width=\linewidth]{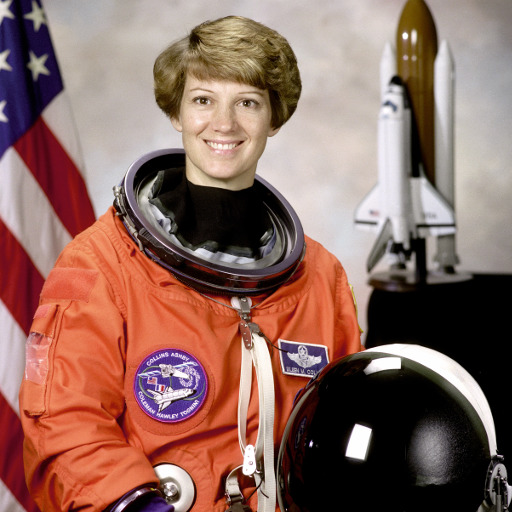}
  \caption{512x512 test astronaut image}\label{fig:astro}
\endminipage\hfill
\minipage{0.32\textwidth}
  \includegraphics[width=\linewidth]{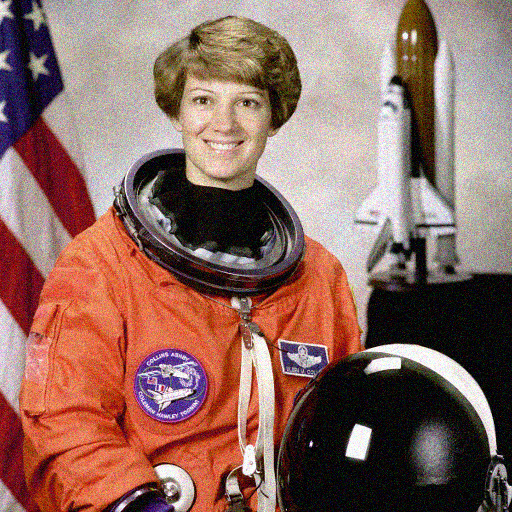}
  \caption{Astronaut image with WAGN($\sigma=20$)}\label{fig:astro_noisy}
\endminipage\hfill
\minipage{0.32\textwidth}%
  \includegraphics[width=\linewidth]{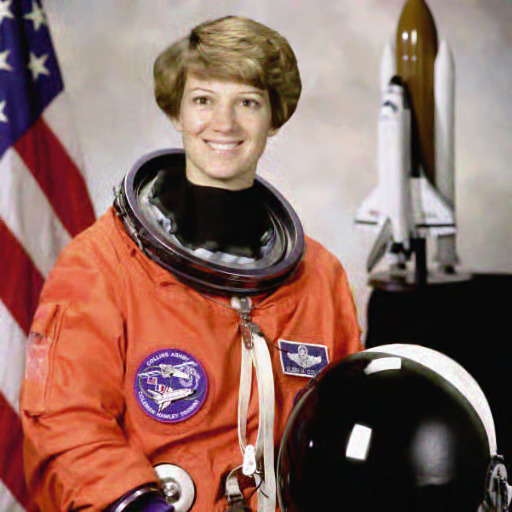}
  \caption{Altered image denoised with modified BM3D}\label{fig:astro_denoised}
\endminipage
\end{figure}

Figure \ref{fig:astro_denoised} shows the output of our modified BM3D algorithm, that takes as an input a previously saved Block-matching table, running on a test image (Figure \ref{fig:astro}) with artificially added White Additive Gaussian Noise of $\sigma = 20$ (Figure \ref{fig:astro_noisy}).

This was done to have a full system verification, but due to lack of time, we didn't implement the Pick N Best stage. The denoising results of our implementation will depend on the same parameters cited by Lebrun \cite{lebrun}, as we are accelerating the same computations he did for the Block-matching.

\subsection{Verification}
We verified the validity of our architecture on simulation by using a minimal reference image with predictable results. This image is build such as each row equals:
$I(i, j) = j$

The image has the following properties:
\begin{itemize}
    \item The sum is a function of the offset and is constant
    \item On each row it's a multiple of the row till the full block is within the frame (see Fig. \ref{blockchallenge})
    \item On each column it grows by same step once the full block is within the frame (see Fig. \ref{blockchallenge})
\end{itemize}

This enabled us to detect and correct architectural misalignment with the simulation wave analysis as every edge case has strong markers, intermediate results can be interpolated and edge locations are identifiable. 

We can also save the simulation output which is the list of final sum tables, but depending on the input size it can take up a very long time, up to 4 hours for a single 360p image. This wasn't used but could be in the future by passing it as an argument to an open BM3D implementation to have a visually perceptible result.

While we initially planned to use the reference BM3D implementation as verification tool and full system integration, but we ended up missing the Pick N Best step, so instrumentation of BM3D to load pre-computed best N blocks couldn't be used.

Given our current stage, some linear transformations must be made in order to align the indexes of the reference BM3D implementation with ours, such as empty values for total sums outside the frame (our architecture writes-back only the valid sums), and revert the window size exploration (we go from corner to central pixel, reference goes from central pixel to corner).

\subsection{Synthesis}

\begin{figure}[h]
    \centering
    \includegraphics[width=0.5\linewidth]{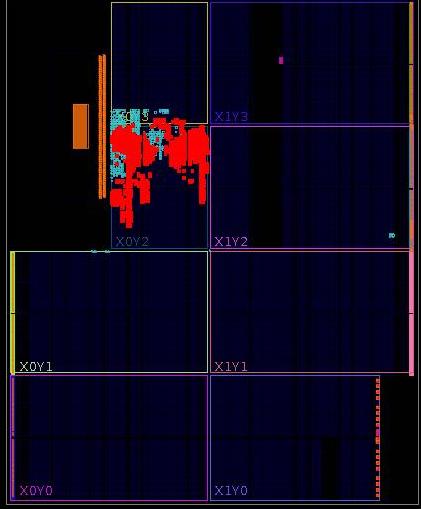}
    \caption{Resource utilization synthesis with parameters 720*1280, search window 32, block size 8, and 4 parallel channels. In red is the full system, blue is overhead to synthetize it.}
    \label{synthesis}
\end{figure}

For synthesis, we wrapped our design with an AXI mapped register file, which gave us the setup to block the tool optimizing for constant wires during synthesis. We used as parameters an image size of 720*1280 (transposed 720p to lower utilization), a window size of 32, and block size of 8.
Our target device was the Zynq 7030. 
Our design was initially able to achieve 100 MHz without any optimizations and 250MHz with the Section 4.2 optimizations while having some margin slack. Previous designs of the BM3D algorithm achieved a maximum frequency of 125 MHz \cite{cardoso}. Although 125 MHz is cited, implementation of FFT's \cite{FFT} and the rest of the algorithm are usually extremely high performing making us believe that the 125 MHz margin comes from the block-matching stage itself.
In terms of resources utilization see Fig \ref{synthesis}, the diff square stage uses 2 BRAM's, each channel worker uses 3 BRAM, the rest of resources are not significant. The total system used less than 6\% BRAM and 2\% of other available resources. While Zynq 7030 is a bigger and bulkier version of the Zynq 7020 which is used in the Apertus Project \cite{Apertus}, timing results are still valid and we use a small fraction of resources no matter the Xilinx module.

\subsection{Theoretical throughput}
Interpolating from our results, running at 250MHz, 16 parallel channels, and if the stride is enough to compensate for the Best N Blocks stage throughput reduction, we can calculate that an image of of 720*1280 can get block matches coordinates in 0.13 seconds while using 30\% of Zynq 7030 BRAM, and less than 10\% of other resources, leaving plenty of space for the rest of BM3D. If the following stages aren't the bottleneck and block-matching is the slowest stage, we could theoretically achieve between 5 and 10 fps for an HD ready image as stages can be pipelined.

\section{Conclusion}
We've managed to expose many of the underlying difficulties and data dependencies of the block-matching step. Our results are promising as we were able to run double the frequency of the previous solutions \cite{cardoso} for the most compute intensive stage, and exploit high amounts of parallelism. We explored an approach that has never been accelerated before, the building differential images instead of computer per window size.

The drawback of the stream-based approach is that we require to store full rows, which are usually between 10x-40x bigger than the window size.
Although the stream-based solution seems worse than the block-based one in terms of scaling with the image size, it could potentially be beneficial for smaller images as there isn't the overhead of synchronization and complex memory patterns of the block-wise approach.

While doing this project we learned a lot in what kind of design patterns can be used both for image processing and any stream based algorithms, such as Stencil Algorithms, Integral Image, prereading and precomputing, buffering, channel synchronization, memory management and more. This knowledge will without doubt be helpful in the future when building any different kind of accelerators.
\newpage
\clearpage
\section{Appendix}

Prerequisites: sbt and scala \\

Running the simulations
\begin{verbatim}
cd RTL
sbt
test:compile
test:testOnly duvs. [TAB] // Here we can see every test
                          // The full system with reference image
                          // is duvs.SteamFullSystemTester

test:runMain duvs. [TAB]  // Here we can see the verilog generators
                          // The full stream based system
                          // is duvs.StreamFullSystem
                          // The standalone synthetizable system 
                          // is duvs.AXIStreamFullSystem
                          // The AXI is functionally incorrect but
                          // it's only purpose is to ensure
                          // that no signals are optimized
                          // This verilog files can be used by Vivado tools
\end{verbatim}

\bibliographystyle{plain}

\end{document}